# *Vortex phase deterioration* common path interferometry

Pritam P Shetty[1], Hemalatha V[1], Mahalingam Babu[1], Jayachandra Bingi[1, *]

[1] Bio-inspired Research and Development (BiRD) Laboratory, Photonic Devices and Sensors (PDS) Laboratory, Indian Institute of Information Technology Design and Manufacturing (IIITDM), Kancheepuram, Chennai 600127, India

* Corresponding author, email-id: bingi@iiitdm.ac.in

## Abstract

Common path interferometers (CPI) are significant due to their compactness and vibration resistance. The usual challenge in CPI would arise due to a very small separation between reference and sample beams, where sending a reference beam through a sample is considered as a limitation. But this limitation also makes it difficult to probe the interaction of beams with material as a function of their phase structure. This study can pave the way for a new kind of interferometry that can provide unique phase signatures to study the sample. The paper proposes and demonstrates a novel approach based on thermo-optic refraction, to send both beams through the sample and probe the phase deterioration due to the relative interaction of beams in the material medium. Here, thermo-optic refraction interferometry (TORI) allows the superposition of a higher order vortex beam with a non-vortex beam through the phenomenon of thermal lensing. The non-vortex beam is made to expand in a controlled fashion by another laser. The relative interaction of the expanding non-vortex beam and the vortex beam within the sample, results in the output interferogram. The phase deterioration analysis of the output interferogram elucidate medium driven phase changes. This technique is demonstrated using the milk samples by recording the RMS azimuthal phase deterioration of the OAM beam.

## Introduction

Light beams carrying orbital angular momentum (OAM) has higher interaction with scattering fluid media due to transfer of OAM to the constituent particle [1]–[3]. Studying the OAM beam transmission can give a wealth of information on the constituents for the fluid media of interest[4]–[7]. Generally, phase structure of the optical fields are measured using interferometers where sample and reference beams interfere to give interferograms. Interferograms carry phase information of the sample beam since phase of reference beam is known, and it is not passed through sample[8]. Interferometers like Michelson and Mach-Zehnder have separate paths for the sample and reference beams but are complicated to use and susceptible to external vibration. Common path interferometers(CPI) are a category of interferometers where sample and reference beams travel through same path and hence less susceptible to vibrations. Commonly used CPI for fluid sample analysis is Youngs interferometer but the sample and reference beam separation is small. Hence, microfluidic flow cells are used to pass sample beam through specimen and reference beam unaffected[9], [10]. This limitation of having small sample size in conventional interferometry is arising due to the usage of sample and reference beams with same phase structure. To overcome this, one beam can be replaced with the phase structured OAM beam that gives the relative interaction of vortex and non-vortex beams as output. This information is unique to the medium. Interferometers that use OAM beam on one arm and interact with a medium of interest and interferes with the reference beam is well reported[11]–[13].

The optical kerr effect is a phenomenon where refractive index of medium is changed when an intense beam of light passes through it[14], [15]. Most nano fluids have been reported to exhibit strong optical Kerr effects like self-focussing/de-focussing and spatial self-phase modulation (SSPM). When laser beams undergo SSPM they expand in far-field forming concentric circular rings. The Optical Kerr effect induced by one laser beam used to affect the phase of another beam is called spatial cross phase modulation (SXPM)[16], [17]. In this research, $MoS_2$ nanofluid is used and interference between two optical fields is achieved by SXPM. Hence this effect can introduce the interference between $0^{th}$ and $1^{st}$ order of OAM beams.

Here we propose thermo-optic refraction interferometer (TORI) which easy to setup and allows phase shifting capability. The OAM carried by Laguerre Gaussian(LG) beam causes it to interact more with particles in turbid media than the optical field without OAM[2], [3]. Hence, two beams with and without vortex phase structure experience different scattering effects when passed through turbid media simultaneously. Further, the phase maps extracted from the interference of these beams is used to study relative interaction of beams with milk sample.

## Methods

### Material synthesis and characterization

2D nanoflakes dispersion of $MoS_2$ with average flake thickness of 153.5±23 nm was prepared by liquid phase exfoliation technique. To achieve a stable dispersion, $MoS_2$ nanoflakes were dispersed in 3% w/w Polyvinylpyrrolidone (PVP) polymer solution. The detailed method of $MoS_2$ nanofluid preparation is given in article[18]. UV-Vis spectra of $MoS_2$ nanofluid indicated broadband absorption in visible region(400nm-700nm). To study the transmission of optical vortex through turbid media, milk samples of different concentrations were prepared. Here packaged milk which is pasteurized and toned were used. Toned milk is a combination of buffalo milk and skimmed (fat less) milk. Milk samples of different dilution factors i.e., 1, 1/2, 1/4 and 1/8 were prepared by two-fold serial dilution. Samples were diluted using de-ionised water. Transmission Spectrophotometry (Supplementary Fig. S1) indicated 1.1%, 4%, 18% and 40% transmittance at 630nm wavelength for milk samples of dilution factors 1, 1/2, 1/4 and 1/8 respectively.

# Experimental setup

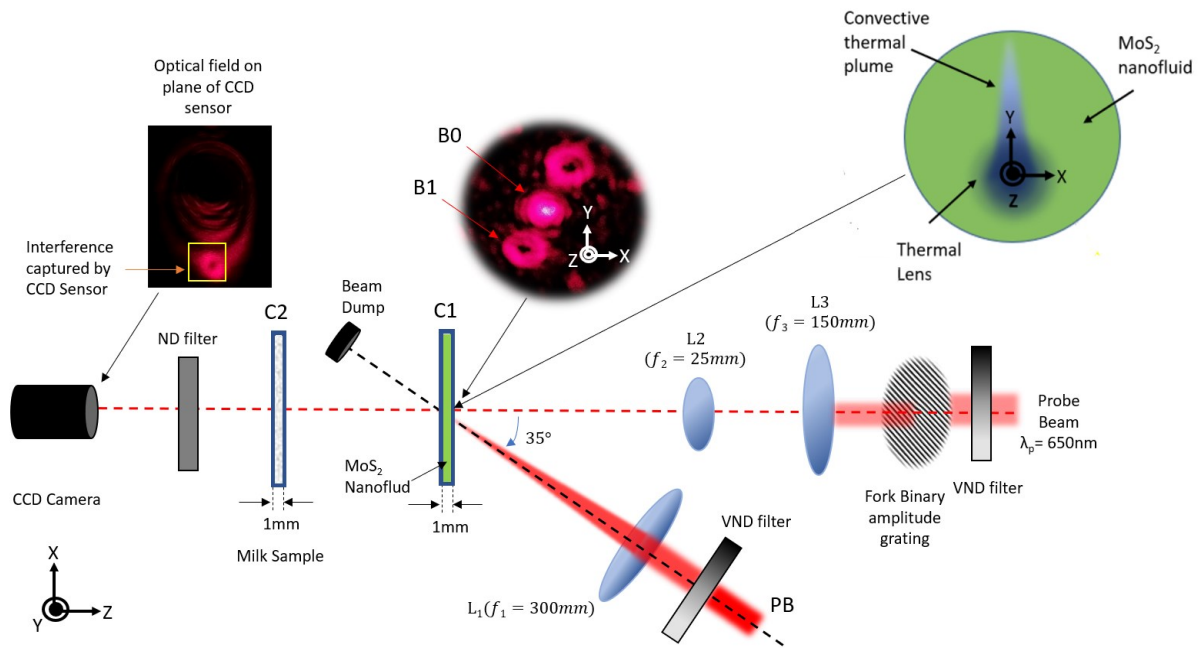

*Figure 1: Experimental setup for Thermo-optic refraction interferometer. PB- Pump beam, B0, B1 are $0^{th}$ and $1^{st}$ order diffraction beams generated by amplitude fork grating, B0 is gaussian beam and B1 is a vortex beam($LG_{0,1}$). C1, C2- glass cuvettes of pathlength 1mm*

Fig. 1 shows optical setup for thermo-optic refraction interferometer. Pump beam (PB) is generated using laser diode of output power 100mW and wavelength 650nm. Beams B0 and B1 are generated using laser diode of output power 50mW and wavelength 650nm. Variable neutral density (VND) filters are used to control beam power of probe and pump beam. A probe beam is passed through binary amplitude fork grating followed by two lenses L2 and L3 which generates Fourier transform of the fork grating pattern and project it. As a result of this, diffraction patterns are formed on the plane of cuvette 1(C1). The diffraction patterns consist of a central $0^{th}$ order diffraction which has gaussian like distribution(B0) and $1^{st}$ order diffraction on either side. The $1^{st}$ order diffraction pattern is a vortex beam(B1) with Topological Charge (TC) corresponding to fork pattern of binary grating. Here, the interference initiated by the grating is responsible for topological charge. Further, topological charge = (Number of fork prongs - 1). The beam powers of B0 and B1 are measured to be ~ 0.3mW and ~0.15mW respectively. Using lens L1, PB is focussed on C1 so as to coincide with the position of beam B0. A thermal lens is formed within $MoS_2$ nano fluid contained in C1 due to focussed PB and its strength is controlled by adjusting power of PB using VND. B0 undergoes SXPM due to thermal lens induced by PB. Both optical fields B0 and B1 pass through milk sample in C2 and then interfere at the plane of CCD camera. Interference fringe contrast can be adjusted by controlling output power of LD2 using VND. A neutral density filter is used to avoid saturation of CCD sensor. The distance between CCD camera and C2 is 32cm.

# Results and discussion

## Thermo-optic refraction interferometer (TORI)

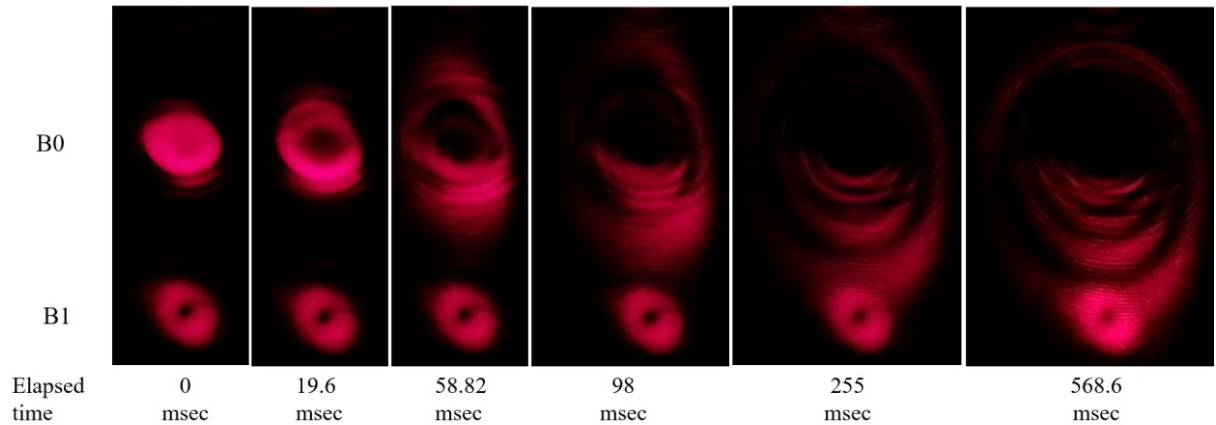

*Figure 2: Time evolution of B0 (0th order diffraction) beam bloom due to thermal lens induced by pump beam*

Fig. 2 shows time evolution of $0^{th}$ order diffraction pattern bloom caused by the thermal lens within MoS$_2$ nanofluid. Beam B0 forms concentric diffraction rings like intensity distribution in far field. This thermal bloom allows interference of beams B0 and B1. It is observed that when $1^{st}$ order diffraction is below $0^{th}$ order diffraction, the convective thermal plume emanating from thermal lens at position of $0^{th}$ order diffraction distorts the optical wavefront of $1^{st}$ order diffraction. This is undesirable and avoided by rotating grating vector of amplitude fork grating by ~20° with vertical.

Considering pump beam (PB) to have a gaussian profile with power $P = 100mW$, beam radius $\omega' = 2 \times 10^{-3}m$, beam spot size $\omega = 119\mu m$ at focal length of lens L$_1$ and radius of curvature of beam R. Then the complex amplitude of PB is given by formula:

$$U(x,y,t,z=0) = \left(\frac{2P}{\pi\omega^2}\right)^{1/2} exp\left(-\frac{x^2+y^2}{\omega^2}\right) exp\left(-ik\frac{x^2+y^2}{2R}\right) \qquad (1)$$

PB focused on the MoS$_2$ nanofluid causes an upward convective flow of hot nanofluid due to change in density. As per literature[19] the convective flow velocity($V_x$) within the fluid is derived with respect to the axis of the Gaussian beam. Further, by considering the balance between the buoyant force of heated nanofluid and viscous force the convective flow velocity ($V_x$) from the centre of PB is given by[20]:

$$V_x = \frac{\beta g [\Delta T]_{max} \pi h^2}{16\mu} \qquad (2)$$

Where $\beta = 1100 \times 10^6 K^{-1}$ is thermal expansion of MoS$_2$ nanofluid, $g = 9.8 m/s^2$ is acceleration due to gravity, $[\Delta T]_{max}$ is a maximum rise in temperature in nanofluid, $h = 1.3 \times 10^{-3}m$ is minimum distance from the centre of the beam to the meniscus, $\mu = 1 \times 10^{-6} m^2/s$ is nanofluid viscosity.

Due to absorption of energy from PB by nanofluid there is a change in temperature distribution in nanofluid. The temperature profile at time '$t$' is given by:

$$\Delta T(x,y,t) = \frac{\alpha P}{\pi \rho c_p}\left(\int_0^t \frac{dt'}{8Dt'+\omega^2} \times exp\left(-2\left(\frac{(x-V_x t')^2+y^2}{8Dt'+\omega^2}\right)\right)\right) \qquad (3)$$

Where $D = K/\rho c_p$ is thermal diffusivity of nanofluid, $K = 0.1769\ W/mK$ is thermal conductivity of nanofluid, $\rho = 789\ Kg/m^3$ is density of nanofluid, $c_p = 2440\ \frac{J}{kg.K}$ is specific heat of nanofluid, $\alpha = 68.097 m^{-1}$ is absorption coefficient of nanofluid at 650nm[18].

Pump beam and beam B0 (Fig. 2 at 0 mSec) are beams with gaussian profile. Both beams coincide with each other at the focal point in the MoS$_2$ nanofluid. This causes the beam B0 to expand in the same fashion as the pump beam, because the refractive index contrast created by the pump beam is also experienced by B0. Assuming the thermally induced refractive index change in nanofluid is constant along the direction of propagation of PB. The spatial distribution of phase shift experienced by beam B0 can be calculated by the formula:

$$\Delta\varphi(x,y,t) = \frac{2\pi L}{\lambda}\left(\frac{dn}{dT}\right)(\Delta T(x,y,t) - \Delta T(0,0,t)) \qquad (4)$$

Where $L = 1 \times 10^{-3}m$ is the pathlength of C1, $\lambda = 650 \times 10^{-9}m$ is wavelength of PB and $\frac{dn}{dT} = 3.66 \times 10^{-5} K^{-1}$ is thermal coefficient of refractive index of nanofluid. Fig. 3 shows calculated phase shift $\Delta\varphi(x,y,t)$ experienced by B0 as it passes through MoS$_2$ nanofluid for the time indicated in Fig. 2 respectively

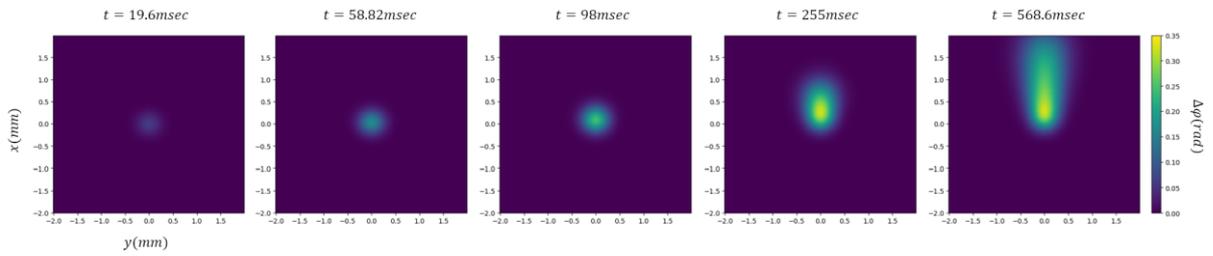

*Figure 3: Spatial distribution of phase shift $\Delta\varphi(x,y,t)$ experienced by B0 as it passes through MoS$_2$ nanofluid for time t=19.6ms, 58.82ms, 98ms, 255ms and 568.6ms*

Furthermore, by controlling the strength of the thermal lens, phase of the interfering B0 can be changed. Therefore, phase shifting interferometry can be performed by controlling power of PB. Fig. 4 shows a typical set of interferograms with two blind phase shifts $\alpha_{21}$ and $\alpha_{31}$ causing fringe shift $\Delta x_1$ and $\Delta x_2$ respectively. Phase shifts up to $\pi$ rad is possible. Blind phase shifting interferometry can be used in metrology applications to generate high resolution phase maps with less noise[21]–[23]. In this research, phase maps are extracted from a single interferogram to study relative scattering of vortex and non-vortex beams through milk samples.

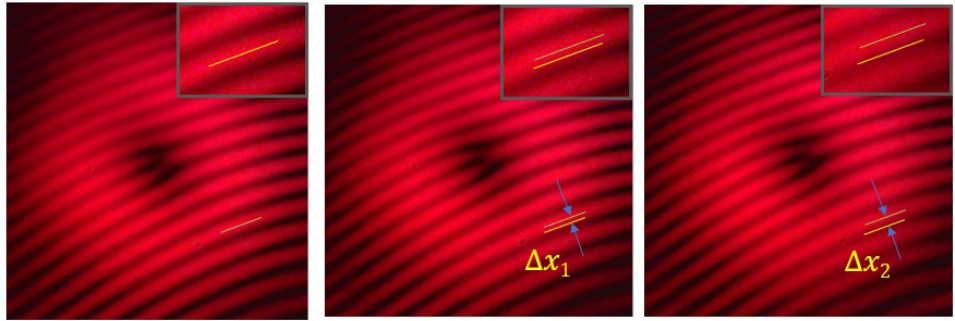

*Figure 4: A typical set of interferograms of optical vortex beam taken by phase shifting of interfering 0th order diffraction field. $\Delta x_1$ and $\Delta x_2$ are fringe shifts associated to arbitrary incremental phase shifts of 0th order diffraction field.*

|  |  | No Sample | 1/8 diluted | 1/4 diluted | 1/2 diluted | Undiluted |
|---|---|---|---|---|---|---|
| TC 1 | Intensity distribution |  |  |  |  |  |
|  | Interferogram |  |  |  |  |  |
| TC 2 | Intensity distribution |  |  |  |  |  |
|  | Interferogram |  |  |  |  |  |
| TC 3 | Intensity distribution |  |  |  |  |  |
|  | Interferogram |  |  |  |  |  |

*Figure 5: Intensity distribution and interferogram of optical vortex with topological charges 1,2 and 3 recorded after transmission through milk samples of different dilution factor 1/8, 1/4, 1/2, 1 and no sample i.e., empty cuvette C2*

Optical vortex beams of TC 1, 2 and 3 are passed through milk samples of different concentrations prepared by 2-fold serial dilution. Beams of TC 1, 2 and 3 are generated using respective binary fork grating. Fig. 5 shows intensity distribution and interferogram of optical vortices after transmission through cuvette 2(C2) with and without milk samples. As the concentration of milk increases, there is higher scattering of the optical field. Consequently, speckle density is proportional to concentration of milk as evident from Fig. 5. Since interferograms are recorded at higher distance (32cm) after transmission through milk samples, there are primarily two types of photons reaching the CCD sensor. They are ballistic photons with high mean free path and slightly scattered snake photons having comparatively lower mean free path while travelling through milk samples[24], [25]. Phase information of all these photons are encoded in the interferograms. This phase information is unique in a way because both beam (B0 and B1) travels through the sample, which is different in comparison to other interferometric methods like Michelson and Mach-Zehnder Interferometers. The phase maps extracted from these interferometers have relative phase change of sample beam with respect to pristine reference beam. Whereas in the case of TORI, Beam B1 carries OAM which can be transferred to the interacting particles[26]. Hence the phase changes in interferograms capture the relative interaction of milk particles with Beam B1 and B0 as a function of its phase structure.

**Phase map extraction from a single interferogram**

Fourier transform method is used to extract wrapped phase maps from a single interferograms as shown in Fig. 5 [27], [28]. Fig. 6 shows Fast Fourier Transform (FFT) spectrum of interferograms taken for optical vortices of TC 1, 2 and 3 after transmission through milk samples of different concentration. FFT spectrum is cropped about the size of filtering window (Fig. 6) to filter out $0^{th}$ order Fourier spectra and include only higher order frequency spectra. Wrapped phase map is extracted from this cropped FFT spectrum as shown in Fig. 7. Patterns of FFT clearly indicate the effect of milk turbidity on the optical field. Frequency spectra becomes less sharp for optical vortices transmitted through milk samples of higher concentration indicating higher scattering and distortion of phase structure.

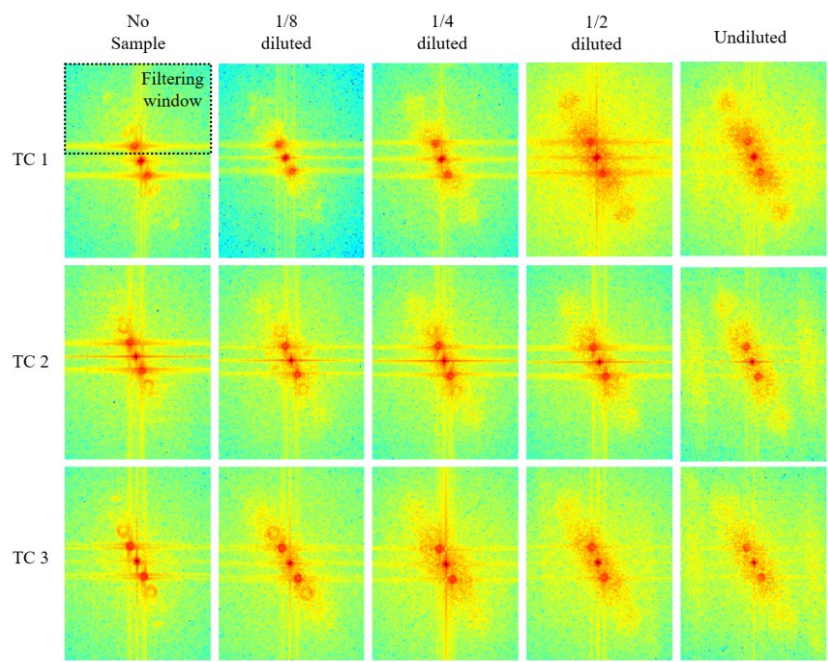

*Figure 6: FFT of single interferogram of optical vortices passing through cuvette C2 without and with milk samples of different concentration.*

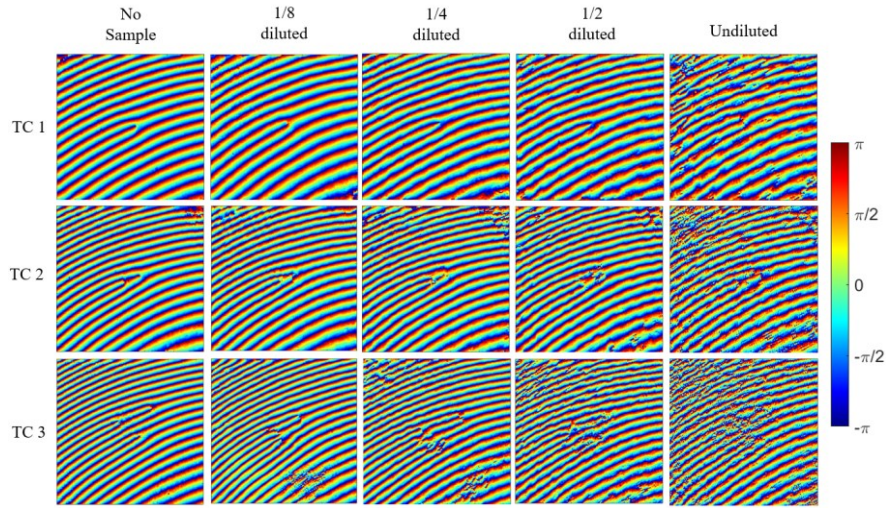

*Figure 7: Wrapped phase maps of optical vortices after transmission through milk samples of different concentration.*

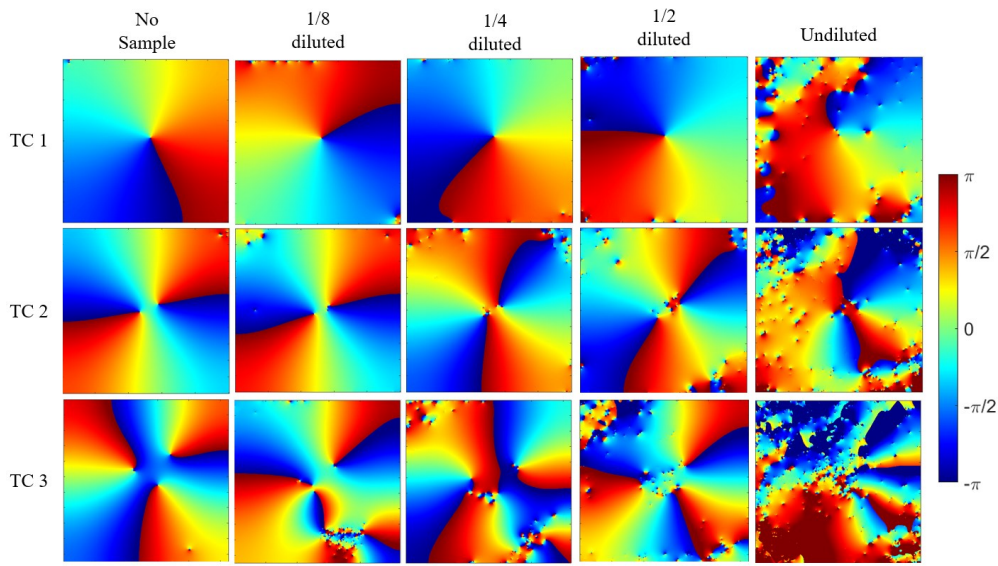

*Figure 8: Unwrapped phase maps of optical vortices after transmission through milk samples of different concentration.*

Now for unwrapping the phase maps given in Fig. 7, an algorithm called iterative transport of intensity equation(TIE) method using the finite difference(FD) and Fast Fourier transform(FFT)is used[29], [30]. Fig. 8 shows unwrapped phase maps of the optical fields after transmission through different milk samples. Unwrapped phase maps indicate that the quantity of vortex phase deterioration depends on the TC it carries. A beam carrying higher TC undergoes higher scattering and its phase structure deteriorates more.

**Vortex Phase deterioration analysis**

Total phase deterioration ($\Delta\varphi$) is calculated by subtracting phase map of optical field after transmission through sample ($\varphi_s$) with the phase map of optical field without sample transmission ($\varphi_n$). Beam centres in both phase maps $\varphi_s$ and $\varphi_n$ should be aligned before subtraction. Then we get

$$\Delta\varphi = \varphi_s - \varphi_n. \tag{5}$$

Laguerre Gaussian beam has spiral phase structure and hence we calculate root mean square (RMS) azimuthal phase deterioration ($\Delta\varphi_{az}$) along the radial direction of the beam. Consider a circle of radius 'r' with its centre over vortex beam axis on phase map $\Delta\varphi$. If $\Delta\varphi_{(\theta,r)}$ is a set of all phase values along the circumference of this circle i.e., azimuthally ($\theta$) from 0 to $2\pi$ and $N_r$ is a set of all data points along the circumference of this circle. Then $\Delta\varphi_{az}$ at distance 'r' from centre of vortex beam is computed from phase map $\Delta\varphi$ using the formula below:

$$\Delta\varphi_{az} = \sqrt{\frac{1}{N_r} \sum_{\theta=0}^{2\pi} |\Delta\varphi_{(\theta,r)}|^2} \qquad (6)$$

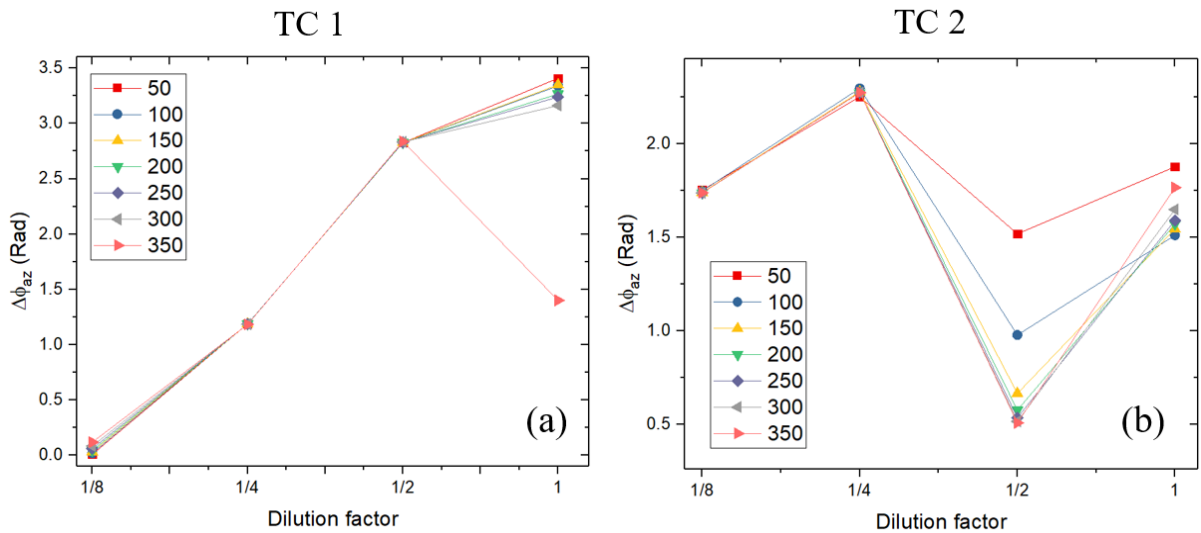

Figure 9: RMS azimuthal phase deterioration ($\Delta\varphi_{az}$) of optical vortex for different dilution factor of milk calculated along radial distance(r) of beam in pixels

Table 1: Average RMS azimuthal phase deterioration ($\Delta\varphi_{az}$) across beam profile for different concentration of milk

| Milk Dilution factor | Fat Concentration (%) | Average RMS azimuthal phase deterioration ($\Delta\varphi_{az}$) | |
|---|---|---|---|
| | | TC1 | TC2 |
| 1/8 | 0.375 | 0.05127 | 1.741433 |
| 1/4 | 0.75 | 1.185611 | 2.274211 |
| 1/2 | 1.5 | 2.829556 | 0.75589 |
| 1 | 3 | 3.024164 | 1.643979 |

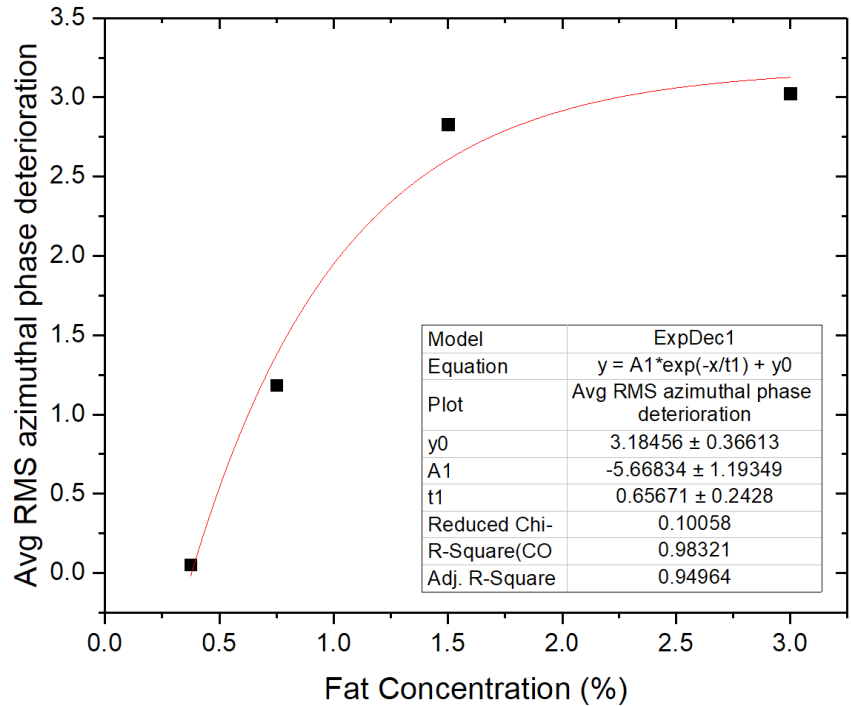

*Figure 10: Calibration curve for estimation of fat concentration for TC1*

Fig. 9 shows azimuthal phase deterioration with increase in concentration of milk for different radial distances of beam in pixels. Here dispersion in data points (at a particular concentration) indicates how phase deterioration is varying across the beam radially. For TC1, the average $\Delta\varphi_{az}$ of overall phase profile is increasing with concentration as seen in Table 1. In the vortex beam the photons near the singularity and away from it are exhibiting different dephasing characteristics with respect to concentration of milk. As evident from Fig. 9(a) at lower concentration of milk phase deterioration is low across beam radius. When the concentration is increased the phase deterioration is also increased almost uniformly for different radial distance of beam. which indicates in the case of TC1 photons near the singularity and photons away from singularity within the beam are dephasing in a uniform fashion up to the dilution factor 1/2. At dilution factor 1, the milk medium naturally poses a very high scattering ambience where deterioration of near singularity photons and the photons away from singularity starts fluctuating. At higher concentration of milk, the vortex region is found to be deteriorating more than the region at beam periphery for TC1.

In case of TC2(Fig. 9(b)), the lower concentration introduced the uniform deterioration along radius of beam but at relatively high value in comparison to TC1. This can be attributed to the increased divergence of TC2. When the concentration is increased, at the dilution factor 1/2 the same differential dephasing is observed between near singularity photons to far Singularity photons, near singularity part of the beam is dephasing more in comparison to far singularity part of beam. In other words, near singularity part of the beam is impacted more by the medium. In comparison to far singularity part of the beam. Once this dilution factor is crossed the high divergence and the high scattering condition completely deteriorate the beam and hence eventually show no trend.

In the case of TC3 we observed, the complete vortex phase deterioration even at low concentration of 1/8 and no meaningful trend was observed. This observation indicates that the low concentration can be effectively probed using higher topological charge beams as they are more sensitive to differential scattering along transverse beam profile.

Fig. 10 shows the calibration curve for determining fat concentration of milk when ($\Delta\varphi_{az}$) is known for TC1. One must note that the calibration curve changes from one topological charge to another topological charge. Hence in the application point of view one must decide the topological charge and then perform the phase deterioration experiment.

## Conclusion

In this work thermo-optic refraction interferometer is successfully demonstrated. The TORI involves use of thermal lensing phenomena for interfering two optical fields, where one of the optical field is a vortex beam. Phase shifting interferometry can be done using TORI by controlling pump beam power. Phase maps extracted from the interferograms showed that phase structure of higher TC beams undergoes higher deterioration. RMS azimuthal phase deterioration indicated that beams undergo different level of deterioration radially depending on its TC. Hence higher topological charge beams(>2) may be used to probe very low concentration of scatterers as they are more sensitive to phase deterioration. This method can pave the way for a new kind of common path interferometry with relative phase changes of vortex and non-vortex beams.

## Author contributions

**Pritam P Shetty**: Conceptualization, Methodology, Validation, Investigation, Visualization, Writing - Original Draft. **Hemalatha V**: Sample preparation, Writing - Review & Editing **Mahalingam Babu**: $MoS_2$ nanoflakes and nanofluid sample preparation and characterization. **Jayachandra Bingi**: Conceptualization, Methodology, Writing - Review & Editing, Supervision.

## Declaration of Competing Interest

The authors declare that they have no known competing financial interests or personal relationships that could have appeared to influence the work reported in this paper.

## Acknowledgements

Authors acknowledge the funding support from DST India under INT/RUS/RFBR/P-262. TORI is a patent pending technology.